\documentclass{article}
\usepackage{epsfig}

\def\abstract#1{{
          \centering{\begin{minipage}{4.5in}\footnotesize\baselineskip=10pt
          \parindent=0pt #1\par 
          \end{minipage}}\par}}

\newcommand{\ket}[1]{|#1\rangle}
\newcommand{\bra}[1]{\langle#1|}

\newcommand{\0}{|0\rangle}
\newcommand{\1}{|1\rangle}

\newcommand{\al}{\ket{\alpha}}
\newcommand{\alm}{\ket{-\alpha}}

\newcommand{\qubit}{\mu\ket{0}+\nu\ket{1}}

\newcommand{\braket}[2]{\langle#1|#2\rangle}

\begin{document}

\centerline{\bf Quantum multiplexing with optical coherent states}
\vspace*{0.37truein}
\centerline{Juan Carlos Garc\'ia-Escart\'in}
\vspace*{0.015truein}
\centerline{\footnotesize\it e-mail: juagar@tel.uva.es}
\vspace*{5pt}
\centerline{Pedro Chamorro-Posada}
\vspace*{0.015truein}
\centerline{\footnotesize\it  Dep. Teor\'ia de la Se\~{n}al y Comunicaciones e Ingener\'ia Telem\'atica.}
\centerline{\footnotesize\it  E.T.S.I. Telecomunicaci\'on, Campus Miguel Delibes, Camino del Cementerio s/n,}
\baselineskip=10pt
\centerline{\footnotesize\it  47011 Valladolid, Spain.}
\vspace*{0.225truein}

\abstract{
In this paper, we propose a novel quantum multiple access technique based on optical coherent states. The information of several coherent state optical qubits is combined into a single qudit, which is the superposition of almost orthogonal coherent states. The original information is encoded into a new Hilbert space with the help of a quantum multiplexer (QMUX) and then recovered at the other end with a quantum demultiplexer (QDEMUX). We introduce the optical tools that complete the coherence state quantum computation model and give the desired circuits. The proposed system can admit a number of users above the optimal limit at the cost of a degradation of the transmitted data. In this and some other aspects, it can be regarded as a quantum analogue of classical Code Division Multiple Access techniques.
}

\section{Introduction}
\label{intro}
Quantum information promises secure communications, faster algorithms and new ways to think about computation and physics \cite{NC00}. The recent developments in quantum computation and communication have reached a point where architectural problems need to be taken into account. As quantum processing units and networks grow in complexity, it is increasingly important to give efficient ways to connect the different elements inside quantum processors and to be able to accommodate multiple users in the same quantum network. 

One particularly important problem is how a finite amount of resources is assigned to a number of competing users. In most communication systems, many different users share the available channel capacity and a series of multiple access techniques are employed. Inspired by one of these techniques, Code Division Multiple Access, CDMA, we will put forward a coherent state quantum multiplexer. The proposed scheme is easily scaled and the only limit is given by the desired transmission quality. Coherent states allow to reproduce some of the most interesting features of CDMA, like the use of almost orthogonal signals. 

The given multiplexing architecture will be an extension of the ideas of optical coherent state quantum computation. We will present the building blocks needed in such an extension and suggest simple physical implementations for them. Finally, an analysis of the efficiency is presented and future possible lines of work are commented.

Section \ref{multiple} introduces the problem of multiple access and gives a brief review of the most extended techniques both in classical and quantum communication systems. One of them, CDMA, is covered in particular detail as the inspiration for the proposed multiple access scheme. Section \ref{QCoherent} reviews the fundamental aspects of quantum computation with coherent states and why it is an adequate representation for a multiple access system. Section \ref{tools} presents the optical elements that are the basic building blocks of our proposal. Section \ref{cohmux} gives a description of the multiplexing system and presents the block diagrams of the quantum multiplexer and demultiplexer. It also discusses how the number of users can be dynamically increased. The section is completed with an example of operation. Section \ref{Conclusion} concludes with a review of the obtained results and a hint of future possibilities. 

\section{Multiple users in communications}
\label{multiple}
Channel capacity is a limited resource. In many situations, the same transmission resources need to be shared by more than one user. \emph{Multiple access} methods are concerned with how $N$ users can share 1 channel. This scenario arises naturally in many classical communication systems \cite{Skl83,Skl01}. For instance, there is only a limited amount of frequencies that can be used to transmit efficiently through the atmosphere and usually limited frequency bands are allocated to each service. Additionally, each band must be offered to users that have to make a new division. In those cases, Frequency Division Multiple Access, FDMA, schemes are applied to coordinate the division of the spectrum. Another usual situation is the one of optical networks, where optical fibre deployment is costly and physical links are scarce. There are different techniques to accommodate many users into the same fibre link. Users can transmit in turns, one user at a time, as in Time Division Multiple Access, TDMA. It is also possible to associate each user to a different wavelength, so that different data do not interfere, as in Wavelength Division Multiple Access, WDMA. These are some of the multiple options that are in widespread use in current communication systems. 

\subsection{CDMA}
\label{CDMA}
\begin{figure}[!ht]
\centering
\includegraphics[scale=0.5]{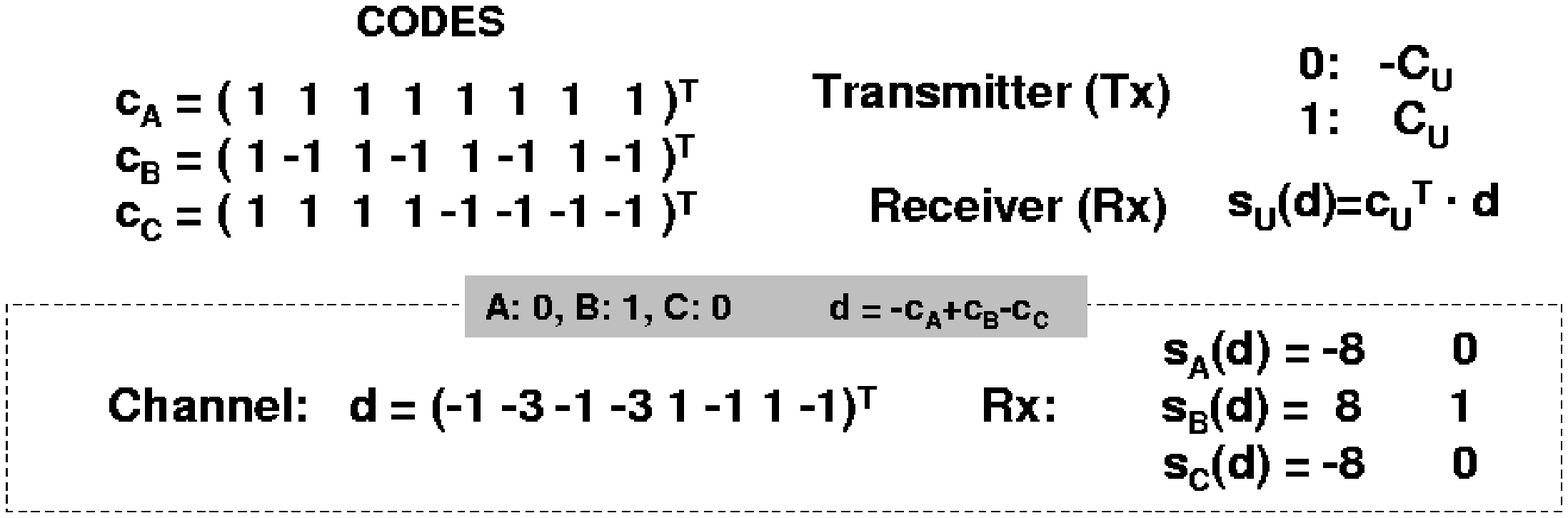} 
\caption{{\bf Example of CDMA multiple access:} CDMA for three users A, B and C that transmit 0, 1 and 0, respectively. At the receiver, 1s will result in a positive weight, 8, equal to the length of the code, and 0s will produce the corresponding negative result, -8. The given codes, $c_A$, $c_B$ and $c_C$ are orthogonal and perfect symbol recovery is possible in an ideal transmission.\label{CDMA1}} 
\end{figure}

One multiple access alternative that has been especially successful in modern wireless networks is Code Division Multiple Access, CDMA \cite{PSM82,Sta01}. In this form of multiple access, all users emit at the same time and their signals spread through all the available bandwidth. In order to avoid interference, the system takes advantage of the algebraic properties of the transmitted signals, which are encoded in orthogonal codes. In CDMA, each user, $U$, is assigned a code $c_U$. User $U$ will transmit the code $-c_U$ to send a 0 and $c_U$ to send 1. The resulting signal at the receiver, $d$, will be the linear combination of all the transmitted data. The receiver can select a channel by applying the corresponding decoding function $s_U(d)$, which is calculated taking the scalar product of $d$ by the code $c_U$ of the user the receiver is listening to. The result of $s_U(d)$ is then compared to a certain threshold to recover the data. If all the users have orthogonal codes, the original 0s and 1s are perfectly recovered (see the example of Figure \ref{CDMA1}).  

\begin{figure}[!ht]
\centering
\includegraphics[scale=0.5]{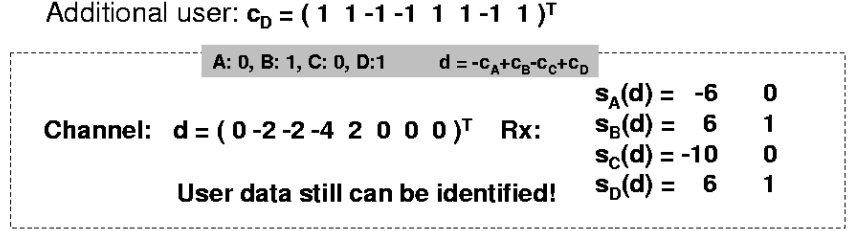} 
\caption{{\bf Example of CDMA with non-orthogonal codes:} A new user, D, is added to the transmission scenario of Figure \ref{CDMA1}. Its code, $c_D$, is not orthogonal to the previous ones, but has a small overlap with them. In this case, where D is sending a 1, decoding the received superposed signal, $d$, still gives weights that can be easily recognized as 1s, the positive values, or as 0s, the negative ones.\label{CDMA2}} 
\end{figure}

Interestingly, the scheme is still valid for almost orthogonal codes (see Figure \ref{CDMA2}). If two user codes have a small scalar product, the decoding process can still be valid if there is low noise and all the signals arrive with the same power. As the number of non-orthogonal codes grows, it becomes increasingly difficult to decode the information of each channel correctly. Similarly, the smaller the overlap between codes, the better the quality of the transmission is. 

The ability to use non-orthogonal codes provides CDMA systems a unique characteristic. Unlike other multiple access schemes, users can be added above the perfect operation limit. For instance, in an FDMA system, if all the spectrum divisions are taken, new users are not allowed to join the channel. Already established connections are given preference and newcomers must wait until one frequency band is liberated. In CDMA, when the orthogonal codes are all assigned, new users can transmit using new non-orthogonal codes, often chosen so as to minimize interference. The quality of the communication will decrease proportionally to the number of the new users. This degradation will be equally suffered by all the participants. As a result, the negative consequences of channel overloading are diffused through all the users and no connection is refused. This gentle degradation of transmission quality can be useful when the number of users is only slightly over the limit. In that case, users do not see an appreciable decrease in the system performance. The system can absorb temporary peaks at a lower efficiency and still treat all the users equally. If the number of users continues increasing, it will eventually cause the system to stop working. The same mechanisms make it robust against noise that is picked up during the transmission.

This is the behaviour we would like to reproduce with our quantum system.

\subsection{Quantum Multiple Access}
\label{HDMA}
Multiple access techniques can also be applied to quantum information. Some of the existing classical infrastructure has already been used for the transmission of quantum data coming from different users. The wavelength division multiplexing experiments \cite{BBG03} and the proposal for FDMA of microwave frequency qubits \cite{OC06} are only some examples of how quantum data can be multiplexed with adapted classical architectures. 

The concepts behind CDMA can be generalized to the quantum case, including intrinsically quantum aspects like entanglement, correlation erasure and superpositions. Starting from basic state swapping concepts, generic quantum multiplexer, QMUX, and demultiplexer, QDEMUX, circuits can be derived \cite{GC07}. 

These circuits convert between the two usual quantum information units, qubits and qudits. Like in the classical communication and computation domains, the most important results and implementations of quantum information processing are based on digital systems, with a discrete set of possible values. The basic quantum information unit, analogue to the classical bit, the qubit, is described as a two-level system that can be in any superposition $\ket{\psi}=\qubit$. $\mu$ and $\nu$ are the complex probability amplitudes associated to each value. A measurement on the qubit basis will give $\0$ and $\1$ with probabilities $|\mu|^2$ and $|\nu|^2$ respectively. These probabilities must sum to 1. An alternative encoding uses multilevelled systems. A qudit is a d-levelled system that can be in any superposition of $d$ possible values
\begin{equation}
\sum_{i=0}^{d-1}\mu_i \ket{i}^d,
 \end{equation}
where the superscript in $\ket{i}^d$ gives the dimensionality of the qudit space. Again, when a measurement is taken, the probability of finding $\ket{i}^d$ is determined by $|\mu_i|^2$ and 
\begin{equation}
\sum_{i=0}^{d-1}|\mu_i|^2 =1.
 \end{equation}
Qubits are a particular case of qudits where $d=2$. 

\begin{figure}[!ht]
\centering
\includegraphics[scale=0.6]{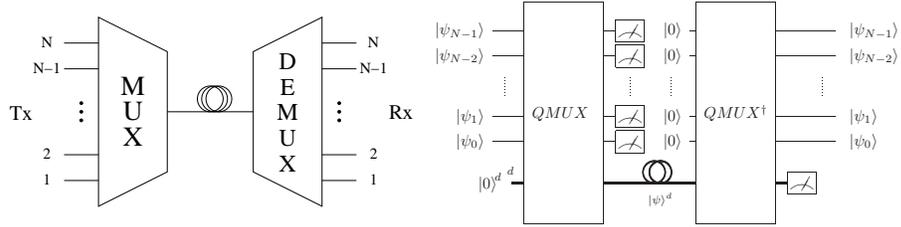} 
\caption{Classical and quantum multiplexing.\label{MUXes}} 
\end{figure}

The QMUX circuit takes $N$ qubits and expresses them in a single qudit of dimension $d=2^N$ that lives in a Hilbert space larger than those of any of the individual qubits. It is this qudit which is sent. The demultiplexer will perform the inverse operation to recover the user information. The architecture adapts the generic classical model with due attention to entanglement and correlation erasure. Deleting information is not trivial in the quantum world and some precautions need to be taken before disposing of the original $N$ qubits. Figure \ref{MUXes} contrasts the classical and quantum multiplexing blocks.

In the QMUX model, the spare qubits are taken to a known state, considered here to be $\ket{0\ldots0}$ without a loss of generality, which is then measured without affecting the qudit. These steps are required to avoid unwanted state reductions. The same known states can be generated later at the receiver to apply the inverse gate and recover the initial qubits at the other side of the channel.

\section{Quantum computation with optical coherent states}
\label{QCoherent}
\subsection{Optical coherent states}
The study of the quantum states of light has been an active field of research since the second half the twentieth century. Optical coherent states were among the first quantum states to be described \cite{Gla63} and have been extensively treated ever since. 

\subsubsection{Description and properties}
Some of the properties of coherent states will be important for our application. In particular \cite{Lou00}:

\begin{itemize}
\item{\emph{Definition:}} Coherent states can be defined by a complex number, $\alpha$, which determines the probability amplitude coefficients of the state when written as a superposition of photon number states. 
\begin{equation}
\label{cohnum}
\ket{\alpha}=e^{-\frac{1}{2}|\alpha|^2}\sum_{n=0}^{\infty}\frac{\alpha^n}{\sqrt{n!}}\ket{n}.
\end{equation}
\item{\emph{Phase and quadrature operators:}} An illustrative representation of coherent states can be made in terms of the complex arguments of $\alpha$. For $\alpha=|\alpha|e^{i\varphi_{\alpha}}$, we can speak of an amplitude $|\alpha|$ and a $\varphi_{\alpha}$ phase similar to those of a classical field.

Coherent states can be represented on the complex XY plane, where X and Y are the real and imaginary parts of $\alpha$ respectively. The real part is in phase with the reference phase and the Y component represents the quadrature part of the signal. Homodyne detection allows to measure these operators by adjusting the phase of a reference local oscillator. 

\begin{figure}[!ht]
\centering
\includegraphics[scale=1]{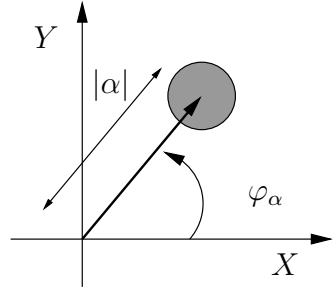} 
\caption{Representation of the $\al$ coherent state in the XY complex plane.\label{coherent}} 
\end{figure} 

Figure \ref{coherent} shows the quadrature representation of the $\al$ coherent state. The state is represented as a circle due to the inherent uncertainty of quadrature measurements. The points inside the circle correspond to measurement results within the standard deviation. The centre of the circle is given by the mean values for each measurement. The measurement statistics for the $\hat{X}$ and $\hat{Y}$ observables are given by the well-known relations:
\begin{eqnarray}
\bra{\alpha}\hat{X}\al&=&Re( \alpha)=|\alpha|\cos{\varphi_{\alpha}}.\\
\bra{\alpha}\hat{Y}\al&=&Im( \alpha)=|\alpha|\sin{\varphi_{\alpha}}.\\
(\Delta X)^2&=&(\Delta Y)^2=\frac{1}{4}.
\end{eqnarray}
The uncertainty is constant for all coherent states, which are represented with circles of the same diameter. 

\item{\emph{Overlap:}} Coherent states form an overcomplete set. Different coherent states are not completely orthogonal to one another, but, for a large enough distance in the XY plane, their overlap can be arbitrarily small. The overlap between two coherent states is 
\begin{equation}
\braket{\alpha}{\beta}=e^{-\frac{1}{2}|\alpha|^2-\frac{1}{2}|\beta|^2+\alpha^{*}\beta},
\end{equation}
and the probability of taking a coherent state $\al$ for another state $\ket{\beta}$ is related to the value of
\begin{equation}
|\braket{\alpha}{\beta}|^2=e^{-|\alpha-\beta|^2}.
\label{overl}
\end{equation}
\end{itemize}

\subsubsection{Why coherent states?}
There are a variety of reason why optical coherent states result attractive for quantum information:
\begin{itemize}
\item They are relatively \emph{easy to produce} with single-mode lasers. In fact, many optical quantum computing proposals approximate number states by a coherent state.
\item There exist efficient \emph{experimental techniques} to manipulate coherent states and to perform X and Y measurement (like homodyne detection \cite{CLG87}). Standard optical elements such as beamsplitters and photodetectors allow for a wide range of operations.
\item There are \emph{many available almost orthogonal states} to choose from. There are infinitely many coherent states, although there will be practical restrictions like noise and power considerations that will limit the usable set. To that respect, coherent states can be related to CDMA codes.
\item They have \emph{well-known properties} and a clear interpretation. They also show quantum behaviour while retaining a classical flavour that simplifies the study.
\end{itemize}

\subsection{Coherent states as qubits}
Coherent states can be used for universal quantum computation \cite{JK02,RGM03}. We will follow the ingenious model of \cite{RGM03} all through the paper, assuming perfect operation of all the gates described there. We will only treat superficially the implementation details, referring the interested reader to the literature on the topic. In this Section, we review the more important aspects that will be essential for our proposal and assume all the other gates and procedures of the model work perfectly.

\begin{figure}[!ht]
\centering
\includegraphics[scale=1]{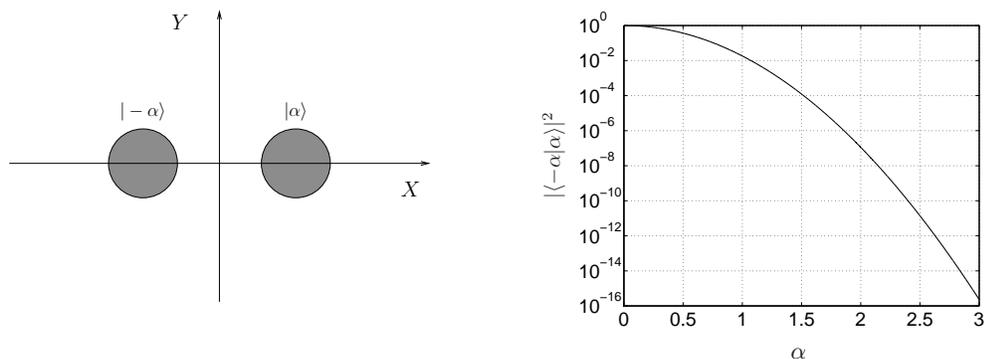} 
\caption{Qubit representation in the XY plane (left) and the evolution of the overlap between logical $\0_L$ and $\1_L$ as a function of $\alpha$ (right).\label{cohqubit}} 
\end{figure}

In this version of coherent state quantum computation, logical states are encoded into coherent states so that $\0_L\equiv\alm$ and $\1_L\equiv\al$, with $\alpha$ being real. Although $\0_L$ and $\1_L$ are not orthogonal, the probability of a wrong identification can be made arbitrarily close to zero. Figure \ref{cohqubit} shows the qubit representation in the XY plane and the values of the overlap for different choices of $\alpha$, as described by Equation (\ref{overl}). Qubits can then be distinguished by homodyne in-phase detection. A negative X measurement will indicate $\0_L$ and a positive result can be interpreted as a $\1_L$ state. This classification will be correct with a high probability and errors will be more unlikely as $\alpha$ grows. This symbol detection is very similar to the 0 and 1 discrimination we saw for CDMA in Section \ref{CDMA}.

Even for relatively small values of $\alpha$, the probability of confusion is so small that can be neglected and will probably not be the dominant source of error when compared with the effects of implementation imperfections or decoherence. 

The available operations on coherent state qubits have also been shown to be able to produce a universal set of quantum gates \cite{RGM03}, i.e. a set of operations that can be combined to produce any desired quantum computation. In many cases, these gates involve going out of the qubit space.  

\subsection{Teleportation}
\label{tele}
Teleportation is the basic operation in coherent state quantum computation. Many of the coherent state logic gates are based on variations of the basic scheme which can teleport a transformed version of the original qubit. Quantum teleportation \cite{BBC93} is one of the basic elements of quantum computation. It allows to reproduce the state of an arbitrary qubit at another location by means of shared entangled states and Bell measurements. 

\begin{figure}[!ht]
\centering
\includegraphics[scale=0.6]{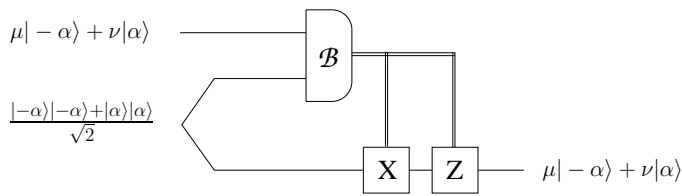} 
\caption{Teleportation of a coherent state qubit.\label{teleportation}} 
\end{figure}

Figure \ref{teleportation} depicts the basic coherent qubit teleportation configuration. At the input we have the qubit that is to be teleported and a Bell coherent state $\frac{\alm\alm+\al\al}{\sqrt{2}}$\footnote{Strictly, this state is not a unit vector. Its components are not completely orthogonal and the cross terms will not be zero. Therefore, the normalization factor is not exactly $\sqrt{2}$. Even so, it will be a good approximation for our range of values of $\alpha$ and a $\sqrt{2}$ factor will be used in the rest of the paper.}.

In the scheme, Bell measurement of the joint state of the qubit and the first state of the Bell pair allows teleportation up to single qubit operations that can be later applied depending on the measured values. The X operation only requires a $\pi$ phase shifter. The classically controlled Z correction is achieved by repeated teleportation until a second sign change cancels the first one, for a probability of one half at each attempt. The coherent Bell basis is not a complete set and there will be a residual small probability of failure. In the basic Bell measurement, which is based on mixing the states in a 50\% beamsplitter, this probability of error is of the order of $e^{-\alpha^2}$. Failures are heralded by a null measurement result and are related to the probability of finding zero photons in the two output ports of the beamsplitter.  

Teleportation of coherent states is not only a state transfer mechanism, but can also have an error correction function. Teleportation can take the states back into the qubit space with high probability. Imagine a qubit that has deviated from the qubit space and is in the $\ket{(1\pm\epsilon)\alpha}$ state. Teleportation can take it back to $\al$ with probability $e^{-\frac{|\alpha|^2|\epsilon|^2}{2}}$ \cite{RGM03}. For small deviations, there is a high probability of recovering the original coherent state.

\section{Basic tools}
\label{tools}
In the following Section, we will describe the multiplexer and demultiplexer setups in terms of functional blocks that are independent on the actual implementation. This Section suggests some possible ways to build these blocks. The list is not exhaustive and alternative elements could be used instead.

\subsection{Entangled cat states}
The most important resource in the whole coherent state computation model is a reliable source of entangled coherent states of the different required types. Production of cat states $\frac{\alm+\al}{\sqrt{2}}$ is far from trivial, especially for a high $\alpha$, and a perfect source has not been found yet. Nevertheless, extensive research is being conducted and good enough approximations are already available. Some of the proposals include using cross-phase modulation in dense media \cite{PKH03} or creating cat states from squeezed photons and detectors \cite{JLR05} (see \cite{GV08} for more alternatives). Probabilistic generation methods are also valid. The entangled resources can be generated besides the multiplexing process and be introduced from whichever source succeeds when needed. 

\subsection{Beamsplitters}
The effect of a beamsplitter, BS, on two input coherent states can be described as $\ket{\alpha}\ket{\beta}\longrightarrow \ket{\cos{\theta}\alpha+\sin{\theta}\beta}\ket{\sin{\theta}\alpha-\cos{\theta}\beta}$. The angle $\theta$ determines the transmission and reflection coefficients, which are $\sin{\theta}$ and $\cos{\theta}$, respectively. Reflection on the lower side produces a $\pi$ phase shift. Inside optical fibre systems, optical couplers can replace beamsplitters and perform the same functions. 

Beamsplitters will be the basic element in our divider, multiplier and adder proposals and can be used to transform between the different cat states we need as a resource. From a $\frac{\alm+\al}{\sqrt{2}}$ cat state and the vacuum, we can generate $\frac{\ket{-\cos{\theta}\alpha}\ket{-\sin{\theta}\alpha}+\ket{\cos{\theta}\alpha}\ket{\sin{\theta}\alpha}}{\sqrt{2}}$ cats. The same procedure can be repeated as many times as needed to create cats made of more than two coherent states. 

\subsection{Dividers}
In a first order approximation, a BS with a parameter $\theta=\frac{1}{M} \ll 1$ and an empty input port will induce the evolution $\ket{\alpha}\ket{0}\longrightarrow \ket{\alpha}\ket{\frac{\alpha}{M}}$. We consider that the small error taken in the approximation can be neglected. The first state has, however, an amplitude slightly smaller than $\alpha$, as energy conservation requires.

It must also be remarked that the divider is not a cloning machine. Although the outputs are independent for a single coherent state input, both output ports will become correlated for a general input that is a superposition of coherent states. In order to recover an independent evolution, we will need to introduce later additional elements to disentangle the ports. 

\subsection{Multipliers}
A qubit can be amplified to any desired amplitude by teleportation through $\frac{\ket{-\alpha}\ket{-M\alpha}+\ket{\alpha}\ket{M\alpha}}{\sqrt{2}}$ states, which can be easily produced using a $\frac{1}{M}$ divider on a $\frac{\ket{-M\alpha}+\ket{M\alpha}}{\sqrt{2}}$ cat state. Experimentally, large cat states are more demanding to create, though. 

It is important to notice that, unlike dividers, multipliers are only valid for known states with a given value of $\alpha$. The multiplier block is not a general amplifier for arbitrary superpositions of coherent states. In this case, we are applying a variation on teleportation. The setup can be adapted to multiply qudits, but, the more the possible states, the more complex the teleportation circuit and the necessary resource entangled state will be. 

\subsection{Adders}
\label{adder}
Beamsplitters can also be used to add two coherent states. An adder is a block that produces the transformation $\ket{\alpha'}\ket{\beta'}\rightarrow\ket{\alpha'+\beta'}$. This operation, as we need it, is irreversible. 

We will use a beamsplitter similar to the divider that we will call sum BS. The sum BS has a $\theta=\frac{\pi}{2}-\frac{1}{M^m}$, with $M,m>1$. In the range of values of $m$ and $M$ we are going to use, we can approximate the reflection coefficient of the sum BS as $\frac{1}{M^m}$ and the transmission coefficient as $1-\frac{1}{2M^{2m}}\approx1$, where we neglect all the terms above the first order. As it happened in the divider, this does not imply a violation of the conservation of energy, as the exact calculations show. 

We will add two states of different amplitudes so that $\alpha'>\beta'$. The addition will have two steps. In the first one, we apply an $M^m$ multiplier on the first half $\ket{\alpha'}\ket{\beta'}$ input to produce the state $\ket{M^m\alpha'}\ket{\beta'}$. The result is mixed in a sum BS in the second step. 

For a general input
\begin{equation}
\mu_{00}\ket{-M^m\alpha'}\ket{-\beta'}+\mu_{01}\ket{-M^m\alpha'}\ket{\beta'}+\mu_{10}\ket{M^m\alpha'}\ket{-\beta'}+\mu_{11}\ket{M^m\alpha'}\ket{\beta'}
\end{equation}
the sum BS presents an output
\vspace{1ex}
\begin{center}
\begin{tabular*}{1\textwidth}
     {@{\extracolsep{-1em}}clclcl}
\phantom{aaaaaaa}&$\phantom{+}\mu_{00}$&$\ket{-\alpha'-\beta'}\ket{-M^m\alpha'+M^{-m}\beta'}$&$+\mu_{01}$&$\ket{-\alpha'+\beta'}\ket{-M^m\alpha'-M^{-m}\beta'}$\\
&$+\mu_{10}$&$\ket{\alpha'-\beta'}\ket{M^m\alpha'+M^{-m}\beta'}$&$+\mu_{11}$&$\ket{\alpha'+\beta'}\ket{M^m\alpha'-M^{-m}\beta'}$&,
\end{tabular*}
\end{center}\vspace{-3ex}\begin{equation}\label{addereq}\end{equation}
which already has the desired sum state at the first port. Before proceeding, we need to erase the state of the second port without altering the superposition of the sum states. Erasure will be based on teleportation error correction.

The state of the second port we want to erase is $\ket{\pm M^m\alpha'\pm M^{-m}\beta'}$, which can be written as $\ket{M^m\alpha'(\pm 1\pm \frac{\beta'}{\alpha'}M^{-2m})}$ and taken as a $\ket{\pm M^{m}\alpha'}$ state with an error
\begin{equation}
\label{errorsum}
\epsilon=\pm \frac{\beta'}{\alpha'} M^{-2m}. 
\end{equation}
For the erasure, we mix the second port with the cat state $\frac{\ket{-M^{m}\alpha'}+\ket{M^{m}\alpha'}}{\sqrt{2}}$ using a 50\% beamsplitter and then measure the photon number in each port.

After the 50\% beamsplitter, we obtain coherent states $\ket{\pm \frac{M^m\alpha'}{\sqrt{2}}(2\pm \epsilon)}\ket{\pm \frac{M^m\alpha'}{\sqrt{2}}\epsilon}$ and $\ket{\pm \frac{M^m\alpha'}{\sqrt{2}}\epsilon}\ket{\pm \frac{M^m\alpha'}{\sqrt{2}}(2\pm \epsilon)}$. The error correction happens when we detect zero photons at one of the detectors $D_1$ (up) or $D_2$ (down). If no photon is found, we can fairly assume that we originally had the $\ket{\pm \frac{M^m\alpha'}{\sqrt{2}}\epsilon}$ state. The probability of finding zero photons for the $\ket{\pm \frac{M^m\alpha'}{\sqrt{2}}(2\pm \epsilon)}$ state is negligibly small for our choice of $\alpha$ and $M$. 

The protocol succeeds if we find zero photons in one of the detectors, which happens with probability $e^{-\frac{|M^m\alpha'|^2|\epsilon|^2}{2}}$, similarly to what we had for the teleportation error correction of Section \ref{tele}. In the adder, $\alpha'>\beta'$ and we have a bounded error with $|\epsilon|< M^{-2m}$. In this case, the correction succeeds with a probability of, at least, $e^{-\frac{{\alpha'}^2M^{-2m}}{2}}$, which can be made arbitrarily close to 1 by choosing an appropriate value of $m$.

The measurement can introduce a phase shift between the sum terms. The amplitude of the $\ket{\pm \frac{M^m\alpha'}{\sqrt{2}}(2\pm \epsilon)}$ state has a phase of $0$ or $\pi$, depending on the sign. The detector that counts a number of photons $n_p\ne 0$ will associate a $(-1)^{n_p}$ phase to the sum states correlated to the $\ket{-\frac{M^m\alpha'}{\sqrt{2}}(2\pm \epsilon)}$ terms. The final state can be written, up to a global phase, as 
\begin{equation}
\mu_{00}\ket{-\alpha'-\beta'}+\mu_{01}\ket{-\alpha'+\beta'}+(-1)^{n_p}\mu_{10}\ket{\alpha'-\beta'}+(-1)^{n_p}\mu_{11}\ket{\alpha'+\beta'}.
\end{equation}
If $n_p$ is odd, there will be a relative $\pi$ phase shift between the components. On average, this sign shift will happen half of the times we perform an erasure.

In the multiplexer, the result will be equivalent to a sum in which the first qubit has suffered a Z operation. Z correction can be used with a caveat: the sum state cannot be targeted by the usual coherent state teleportation Z gate lest we destroy the information of the smaller amplitude inputs, which will also carry the user's qubits. However, correlated states can be extracted with a divider with a negligible alteration of the original state. For the values of $\alpha'$ and $\beta'$ we will have at the multiplexer, the extracted state will be dominated by the $\alpha'$ component. We can repeat the erasing procedure for the extracted state using a resource cat state of the amplitude of the leading term. Again, on an odd photon count, we will have a Z operation that corrects the first one. This is achieved with probability one half. The procedure can be repeated as many times as needed until the sum superposition has been corrected. 

The adder we have presented is also valid for inputs that are, in turn, sums from previous adder stages. The results will have a leading term and smaller deviations. The approximations we have taken in the discussion are still correct as long as the leading terms in the lower port are of an amplitude smaller than the smallest term in the upper port.  

\begin{figure}[!ht]
\centering
\includegraphics[scale=0.75]{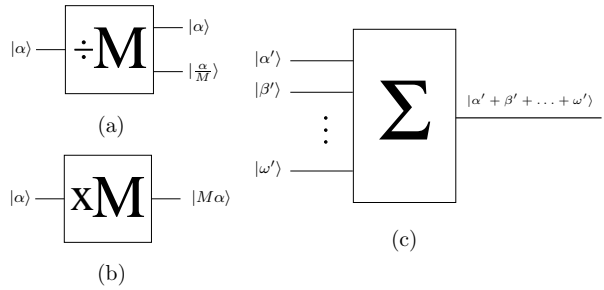} 
\caption{Representation for the divider (a), multiplier (b) and combiner (c) blocks.\label{symbol}} 
\end{figure} 

\subsection{Combiners}
Adders can be cascaded to provide a combiner. We call combiner to any element that merges up to $N$ inputs $\ket{\alpha'}\ket{\beta'}\ldots\ket{\omega'}$ into a single channel with a coherent state $\ket{\alpha'+\beta'+\ldots+\omega'}$, where the constant amplitude correction terms needed for the conservation of energy have been dropped from the discussion for simplicity. Different arrangements inside the combiner will give different correction terms for the input states. We will consider inputs of decreasing amplitudes $\alpha'>\beta'>\ldots>\omega'$, in which we assign an index from $N-1$ to $0$ to the inputs. The $i$-th input will have amplitudes $\pm M^i\alpha$. 

One way to create a combiner is chaining adders in a series configuration. First we need to provide the additional energy that will be lost in the erasure. We can include a premultiplication stage in all the qubits but the last one. A coherent qubit of order $i$, for $i$ from $1$ to $N-1$, will go through a multiplier of factor $M^{i+m}$, with an $m$ that is determined by the sum adder. Now the coherent qubits can be added starting from the qubits of order 0 and 1. Each stage produces a sum state in which the amplitude is no longer multiplied by $M^m$. The higher amplitude terms that have not been added yet still have the $M^m$ factor and enter their adders by the upper port. The series combination will sum the inputs in an increasing amplitude order. The new qubits are incorporated into the sum state one by one, loosing the $M^m$ factor in the process. After $N-1$ adding stages, we will have the desired sum state at the output. 

Combination can also be done in parallel. We can pair the inputs into adders with the same strategy employed in knockout tournaments to determine the matches. Each adder corresponds to one match. This configuration has a maximum number of $\log_2{N}$ stages, with $N-1$ adders \cite{Knu98}. In this case, the sum states that go through more than one adder need to be premultiplied by a factor $M^{N_Am}$, where $N_A$ is the number of adders it enters through the upper port. In each adder, the sum terms of the upper port must be of a higher index than the sum terms in the lower port. Here, we have needed a more restrictive premultiplication because, after the first sum has been performed, our multiplier cannot be applied. The multiplier is not a coherent state amplifier, but specific to the concrete qubits of a determined user of index $i$. It cannot be used on sum states. 

These two groups of cascaded adders form a suitable combiner. The setup can be simplified by an $N\times N$ Multi-Mode Interference, MMI, device with the same transfer matrix as the whole beamsplitter system. MMI devices can give low loss power combiners for any number of input ports \cite{SP95}. The $N-1$ spare ports can be dealt with in the same way as the extra output of the adder scheme. Undesired residual correlation will be erased after the teleportation-like measurement. 

\section{Quantum multiplexing with optical coherent states}
\label{cohmux} 
Using these blocks, we can propose an architecture for quantum multiplexing with many of the characteristics of CDMA. We will show that coherent quantum multiplexers are feasible with existing technology, at least for a reasonable rate of success.

In the following discussion, we will assume that a working teleportation circuit is available with high probability and that we have an unlimited amount of resource Bell states. We will also temporarily forget of any requisite on the maximum value of $\alpha$, assuming any amplitude is possible. This assumption can later be dropped at the cost of a reduction in the global efficiency. Finally, we will usually take first order approximations and establish the order of magnitude of the involved probabilities instead of presenting rigorous, detailed calculations that would obscure the proposal.

\subsection{Multiplexing}
The task of the multiplexer is to combine the information in a way that allows an easy separation at the receiver. The obvious solution is to use a combiner to merge the qubits into one channel. This approach has an important shortcoming: more than one state end in the same multiplexed state in the transmitted qudit. For instance, for the combiner of the previous section, the result of merging two logical $\0_L$ states and a logical $\1_L$ state would result in the same coherent state, $\alm$, irrespective of which qubit contains the $\1_L$. Logical $\ket{100}_L$ would interfere with $\ket{010}_L$ and $\ket{001}_L$, which can be a problem at the reception, where separation should be straightforward. The confusion comes from the irreversibility of the combiner step. A small amount of preprocessing and planning can correct this. In our scheme, distinguishability is achieved by multiplying the coherent states of the qubits by a different factor for every user. Alternatively, users could have different values of $\alpha$ for their qubits from the start. 

\begin{figure}[!ht]
\centering
\includegraphics[scale=0.8]{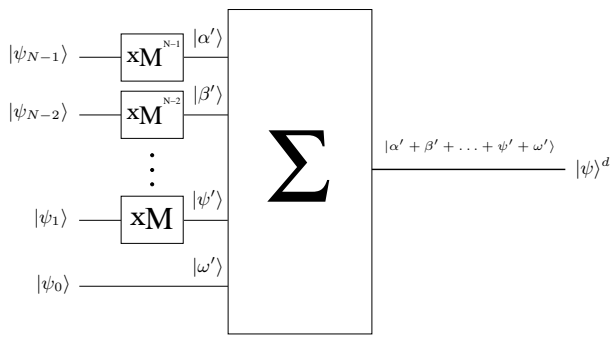} 
\caption{Quantum multiplexer for coherent state qubits.\label{cQMUX}} 
\end{figure} 

Figure \ref{cQMUX} shows the complete architecture of the multiplexer. Each qubit is expressed as $\mu_i\ket{-M^i\alpha}+\nu_i\ket{M^i\alpha}$, where $i$ is the index of the qubit position. Different qubits are encoded at different levels of amplitude and can be later separated. Now, each logical combination belongs to a different state in the qudit. For $M\ge 3$ those states are at least as separated in the XY plane as the original qubits and confusion is highly unlikely. 

The resulting state will be encoded in $d=2^N$ almost orthogonal coherent states. These states define the coherent qudit $\ket{\psi}^{2^N\!}\!$. The superposition can be written with the formula:
\begin{equation}
\ket{\psi}^{2^N}=\sum_{l=0}^{2^N-1}\left( \prod_{j=0}^{N-1} \mu_j^{b_j^l\oplus 1}\nu_j^{b_j^l} \right)\left|\left( \sum_{k=0}^{N-1} (-1)^{b_k^l\oplus 1}M^k\right) \alpha \right\rangle,
\end{equation}
where $b_k^l$ is the bit that describes the term $2^k$ in the binary representation of the number $l$ that is encoded by the logical qubit states and $\oplus$ represents the logical XOR operation.

The probability of success of each addition of the multiplexing process depends on the kind of combiner we use. For a series configuration, the probability of success for the addition of user $i$ is limited by the erasure procedure. We sum states $\alpha'=M^i\alpha$ and $\beta'=M^{i-1}\alpha$. From Equation (\ref{errorsum}), we can see that the error we correct in the erasure is $\epsilon=\pm \frac{\beta'}{\alpha'}M^{-2m}=\pm M^{-2m-1}$. The new user will be added with probability $e^{-\frac{M^{2i-2m-1}\alpha^2}{2}}$. We can adjust the $m$ factor of each adder so that $m\ge i$. In this case, the probability of success is at least as high as the probability of error correction of an $\ket{\alpha}$ state with an error $\epsilon=\frac{1}{M}$, which will be a reference figure in the demultiplexer.

\subsection{Demultiplexing}
At the receiver, the qubits must be extracted. The main difference between quantum and classical transmission is that information cannot be copied. In our scheme, we cannot generate multiple independent copies like in CDMA. This fact, when combined with our encoding, imposes a sequential ordered recovery. We must know in advance the order in which the qubits are going to be extracted.

\begin{figure}[!ht]
\centering
\includegraphics[scale=1]{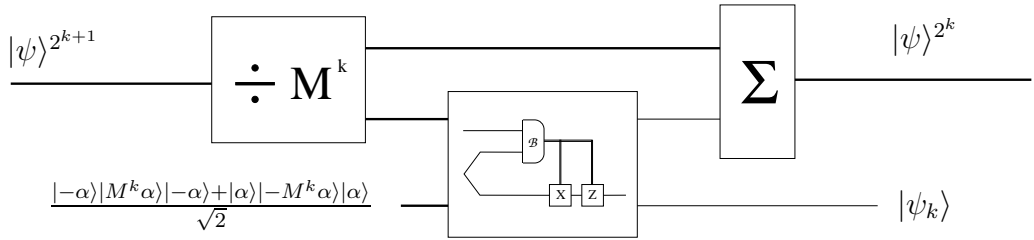} 
\caption{Quantum demultiplexer cell for the $k$-th qubit. The qubit is decoupled from the coherent qudit, which is then put back into the channel.\label{cDEMUX}} 
\end{figure}

Figure \ref{cDEMUX} shows the demultiplexing cell for the $k$-th receiver. The input is a qudit $\ket{\psi}^{2^{k+1}}$ with the information of all the qubits with indices from 0 to k. After the processing, the $k$-th qubit is extracted and only the information of the qubits with indices from 0 to k-1 is kept in the output qudit $\ket{\psi}^{2^k}$. 

The procedure is divided into two stages. In the first one, the coherent state superposition goes through a splitter so that the state is reproduced, divided by $M^k$, in another port. This state is projected to $\ket{\pm\alpha}$ using a teleportation procedure that will also produce an additional output to take care of residual correlations. The first stage ends with three entangled ports, one with the original qudit and, associated to each qudit value, the corresponding coherent state of the $k$-th qubit and an ancillary state that will be used to erase entanglement. The second stage removes all correlation between qubit and qudit.

After the division by $M^k$, the terms of the superposition on the divided port are of the form 
\begin{equation}
\left|\pm\alpha+\frac{\alpha}{M^k}(\pm M^{k-1}\pm{M^{k-2}\pm\ldots\pm1})\right\rangle,
\end{equation}
where the concrete signs depend on the encoded state. Each of these states is still entangled to a different qudit value. Attending to the contents of the extracted state, there are two dominant groups of states. One group corresponds to the logical $\0_L$, with amplitudes close to $-\alpha$ and the other corresponds to logical $\1_ L$, with amplitudes closer to $\alpha$. All of them can be seen as $\ket{\pm (1\pm \epsilon )\alpha}$ states, states in which an error has corrupted the original information. Notice that, in the worst, maximum distance, case 
\begin{equation}
\label{demuxerror}
\epsilon=\frac{1}{M^k}\sum_{i=0}^{k-1}M^i=\frac{1}{M^k}\frac{1-M^k}{1-M}=\frac{1}{M^k(1-M)}-\frac{1}{1-M}.
\end{equation}
If $M\gg1$, $\epsilon$ is of the order $\frac{1}{M}$, which is also small. After the teleportation, all the states within the same group will be projected to the same final state with probability $e^{-\frac{\alpha^2}{2M^2}}$, which is close to 1. The information of the additional qubits has been treated as an error to be corrected. The effects of noise and qudit transmission errors will also be cleaned if they are of a magnitude similar to $M^k\alpha\epsilon$. 

Here, the only difference with respect to basic coherent qubit teleportation is the presence of an output to undo correlations. Teleportation is based on a joint Bell measurement of the incoming state and the first element of the entangled state
\begin{equation} 
\frac{\alm\ket{M^k\alpha}\alm+\al\ket{-M^k\alpha}\al}{\sqrt{2}}. 
\end{equation}
After teleportation, states related to the $\0_L$ of the $k$-th qubit will be entangled to $\ket{M^k\alpha}\alm$ states and qudit states related to the $\1_L$ will be entangled to $\ket{-M^k\alpha}\al$ states. The $\ket{\pm M^k\alpha}$ states will erase the trace of the $k$-th qubit after the combiner with a probability $e^\frac{M^{2k-2m}\alpha^2}{2}$, which, for $m=i+1$, will be similar to the probability of extraction\footnote{In the adder, the $\ket{\pm M^k \alpha}$ states will be multiplied by $M^m$ and added to the multiplexed state. The probability of success can be deduced from Equation (\ref{errorsum}) considering $\alpha'=M^k\alpha$ and $\beta'=M^k(1\pm \epsilon)$, for the $\epsilon$ of Equation (\ref{demuxerror}). In this case, $\frac{\beta'}{\alpha'}$ tends to 1 and the error to be corrected in the erasure is, approximately, $M^{-2k}$.}.

By repeating this procedure, the $n$ qubits can be extracted from the qudit.

\subsection{Scalability and users above the limit}

This scheme can be easily scaled. A new user can be added at any moment just by assigning it a new power of $M$ to multiply its qubits. Theoretically, this could be done for any number of users, but practical limits will set a maximum amplitude for coherent states and a maximum $M^k\alpha$ value that can be reached. 

If there is a limit around $\beta$ on the amplitude, a possible heuristic to accommodate $n$ users would be to choose $M=\left(\frac{\beta}{\alpha}\right)^{\frac{1}{N-1}}$, allowing for the smallest available value of $\epsilon$. If, later, new users must come into the system, they can still be added by defining a new $M'=\sqrt{M}$. This will create new slots to place the qubits of the new users. The price to pay is a general reduction of the efficiency. Now, some of the teleportations will reduce their probability of success from the order of $e^{-\frac{|\alpha|^2}{2M^2}}$ to $e^{-\frac{|\alpha|^2}{2M}}$. 

This behaviour reproduces the gentle degradation in transmission quality of CDMA. In CDMA, almost orthogonal codes created noise that masked the legitimate signal. Here, states are closer in the XY plane and separation is progressively more difficult as new users occupy the new slots.

\subsection{Example} 
To better illustrate the protocol, we can study the multiplexing of three qubits. In this example, the values of $\alpha$ and $M$ are chosen to be round numbers for clarity, which will give coherent amplitudes too large for a real world application. We will discuss later how $\alpha$ and $M$ have to be chosen. In the example, the starting point are three independent users, with coherent states qubits with $\alpha=2$. $M$ is set to be 10.

\subsubsection{Multiplexing:}
First, we multiply the qubits by factors of 10:
\begin{eqnarray}
\nonumber
\ket{\psi_0}=\mu_0\ket{-2}+\nu_0\ket{2}& \stackrel{}{\longrightarrow}&\mu_0\ket{-2}+\nu_0\ket{2}.\\
\nonumber
\ket{\psi_1}=\mu_0\ket{-2}+\nu_0\ket{2}& \stackrel{\times 10}{\longrightarrow}&\mu_1\ket{-20}+\nu_1\ket{20}.\\
\nonumber
\ket{\psi_2}=\mu_0\ket{-2}+\nu_0\ket{2}& \stackrel{\times 10^2}{\longrightarrow}&\mu_2\ket{-200}+\nu_2\ket{200}.
\end{eqnarray}

The first adder of the combiner will sum the qubits of users 0 and 1. If we choose a sum BS with $m=1$, we need to apply a M=10 multiplier to the qubit of user 1. At the input of the sum BS, we will have $\mu_1\ket{-200}+\nu_1\ket{200}$ at the upper port and $\mu_0\ket{-2}+\nu_0\ket{2}$ at the lower port. The output will be
\begin{equation}
\mu_1\mu_0\ket{-22}\ket{-199.8}+\mu_1\nu_0\ket{-18}\ket{-200.2}+\nu_1\mu_0\ket{18}\ket{200.2}+\nu_1\nu_0\ket{22}\ket{199.8}.
\end{equation}
The erasure succeeds with a 98\% probability. The resulting state 
\begin{equation}
\mu_1\mu_0\ket{-22}+\mu_1\nu_0\ket{-18}+\nu_1\mu_0\ket{18}+\nu_1\nu_0\ket{22}
\end{equation}
carries the data of both qubits. 

We can now add the state of user 2. If the maximum amplitude for the coherent states is not restricted, we can directly take a sum BS of $m=2$ to create the state
\vspace{1.5ex}
\begin{center}
\begin{tabular*}{1\textwidth}
     {@{\extracolsep{-0.8em}}lclclc}
$\mu_2\mu_1\mu_0$&$\ket{-222}\ket{-19999.78}$&$+\mu_2\mu_1\nu_0$&$\ket{-218}\ket{-19999.82}$&$+\mu_2\nu_1\mu_0$&$\ket{-182}\ket{-20000.18}$\\
+$\mu_2\nu_1\nu_0$&$\ket{-178}\ket{-20000.22}$&$+\nu_2\mu_1\mu_0$&$\ket{178}\ket{20000.22}$&$+\nu_2\mu_1\nu_0$&$\ket{182}\ket{20000.18}$\\
$+\nu_2\nu_1\mu_0$&$\ket{218}\ket{19999.82}$&$+\nu_2\nu_1\nu_0$&$\ket{222}\ket{-19999.78}$&.&
\end{tabular*}
\end{center}
\vspace{-2ex}\begin{equation}\end{equation}

The probability of success of the erasure is, for the worst case, of almost the 97,61\%. However, we have needed to produce coherent state superpositions of amplitudes of the order of $10^4$. We can reduce the required amplitude if we allow for a higher rate of failure.

For instance, for $m=1.4$, the multiplication factor would be $M^m=10^{1.4}\approx 25.12$. Now, we have at the upper port of the sum BS coherent states of amplitude 5024, an order of magnitude below the amplitudes of the previous case. The probability of success in this case would be around the 68.14\% for the maximum error. 

In both cases we have commented, the qubits are combined into the state
\vspace{1.5ex}
\begin{center}
\begin{tabular}{@{\extracolsep{-0.35em}}lc@{\hspace{0.3em}+\hspace{0.6em}}lc@{\hspace{0.3em}+\hspace{0.6em}}lc@{\hspace{0.3em}+\hspace{0.6em}}lll}
$\mu_2\mu_1\mu_0$&$\ket{-222}$&$\mu_2\mu_1\nu_0$&$\ket{-218}$&$\mu_2\nu_1\mu_0$&$\ket{-182}$&$\mu_2\nu_1\nu_0$&$\ket{-178}$&\\
$+\hspace{0.3ex}\nu_2\mu_1\mu_0$&$\ket{178}$&$\nu_2\mu_1\nu_0$&$\ket{182}$&$\nu_2\nu_1\mu_0$&$\ket{218}$&$\nu_2\nu_1\nu_0$&$\ket{222}$,&
\end{tabular}
\end{center}
\vspace{-2ex}\begin{equation}\end{equation}
which encodes the original tensor product. There is a one-to-one correspondence with the original logical state 
\vspace{1.5ex}
\begin{center}
\begin{tabular}{l@{}c@{ + }l@{}c@{ + }l@{}c@{ + }l@{}cl}
$\mu_2\mu_1\mu_0$&$\ket{000}_L$&$\mu_2\mu_1\nu_0$&$\ket{001}_L$&$\mu_2\nu_1\mu_0$&$\ket{010}_L$&$\mu_2\nu_1\nu_0$&$\ket{011}_L$&\\
$+\nu_2\mu_1\mu_0$&$\ket{100}_L$&$\nu_2\mu_1\nu_0$&$\ket{101}_L$&$\nu_2\nu_1\mu_0$&$\ket{110}_L$&$\nu_2\nu_1\nu_0$&$\ket{111}_L$&.
\end{tabular}
\end{center}
\vspace{-2ex}\begin{equation}\end{equation}

The probability of success of this reencoding depends on the maximum amplitude for which we can keep a coherent state superposition. We can adjust the parameter $m$ of the sum beamsplitters to reduce the maximum required amplitude at the cost of a lower rate of success.

\subsubsection{Demultiplexing:}
At the receiver, we first extract $\ket{\psi_2}$. We start by splitting the coherent states of the superposition with a 100 divider. The resulting state is, approximately,
\vspace{2ex}
\begin{center}
\begin{tabular*}{1\textwidth}
     {@{\extracolsep{-0.0em}}lclclc}
$\mu_2\mu_1\mu_0$&$\ket{-222}\ket{-2.22}$&$+\mu_2\mu_1\nu_0$&$\ket{-218}\ket{-2.18}$&$+\mu_2\nu_1\mu_0$&$\ket{-182}\ket{-1.82}$\\
$+\mu_2\nu_1\nu_0$&$\ket{-178}\ket{-1.78}$&$+\nu_2\mu_1\mu_0$&$\ket{178}\ket{1.78}$&$+\nu_2\mu_1\nu_0$&$\ket{182}\ket{1.82}$\\
$+\nu_2\nu_1\mu_0$&$\ket{218}\ket{2.18}$&$+\nu_2\nu_1\nu_0$&$\ket{222}\ket{2.22}$&.&
\end{tabular*}
\end{center}
\vspace{-1.5ex}\begin{equation}\end{equation}

The second coherent state can be projected into either $\ket{-2}$ or $\ket{2}$ by teleportation with a probability over the 97.60\% for these values. The entangled resource state $\frac{\ket{-2}\ket{200}\ket{-2}+\ket{2}\ket{-200}\ket{2}}{\sqrt{2}}$, will guarantee that, along with the projection, an accompanying $\ket{\pm200}$ state is created. This state is merged with the multiplexed state using a combiner. If we choose a sum BS with $m=3$, the erasure step of the combination succeeds with a probability above the 97.56\%, or above the 67.66\% for a more moderate $m=2.4$. 

After merging the new state with the original channel qudit, we have
\vspace{1ex}
\begin{center}
\begin{tabular}{l@{}c@{ + }l@{}c@{ + }l@{}c@{ + }l@{}cl}
$\mu_2\mu_1\mu_0$&$\ket{-22}\ket{-2}$&$\mu_2\mu_1\nu_0$&$\ket{-18}\ket{-2}$&$\mu_2\nu_1\mu_0$&$\ket{18}\ket{-2}$&$\mu_2\nu_1\nu_0$&$\ket{22}\ket{-2}$&\\
$+\nu_2\mu_1\mu_0$&$\ket{-22}\ket{2}$&$\nu_2\mu_1\nu_0$&$\ket{-18}\ket{2}$&$\nu_2\nu_1\mu_0$&$\ket{18}\ket{2}$&$\nu_2\nu_1\nu_0$&$\ket{22}\ket{2}$.&
\end{tabular} 
\vspace{-4ex}\begin{equation}\end{equation}
\end{center}

\begin{figure}[!ht]
\centering
\includegraphics[scale=0.9]{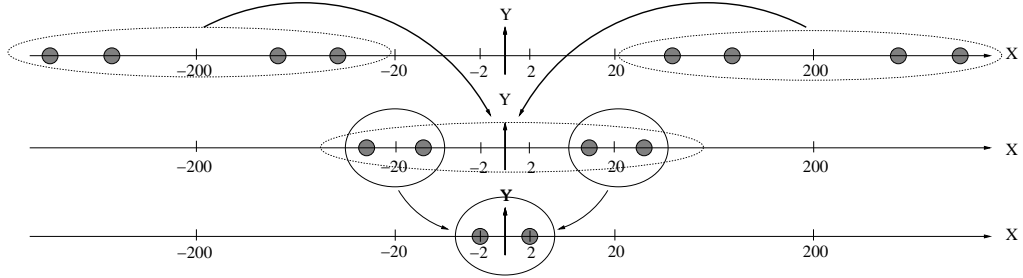} 
\caption[Qubit extraction in coherent state multiplexing.]{Progressive reduction of the qudit space as the individual qubits are extracted. The states that encode the same logical qubit that is extracted at each step are circled by an ellipse.\label{example}} 
\end{figure} 

This state can be factored as
\begin{equation}
(\mu_1\mu_0\ket{-22}+\mu_1\nu_0\ket{-18}+\nu_1\mu_0\ket{18}+\nu_1\nu_0\ket{22})\otimes(\mu_2\ket{-2}+\nu_2\ket{2}),
\end{equation}
where the qubit $\ket{\psi_2}$ has been separated from the superposition.

We can proceed in the same way to extract $\ket{\psi_1}$. This time, we divide by 10. The resulting state is 
\begin{equation}
\mu_1\mu_0\ket{-22}\ket{-2.2}+\mu_1\nu_0\ket{-18}\ket{-1.8}+\nu_1\mu_0\ket{18}\ket{1.8}+\nu_1\nu_0\ket{22}\ket{2.2}. 
\end{equation}
Teleportation will correct deviations from $\ket{\pm2}$ with a probability slightly above 98\%. 

After the projection and combination of the $\ket{\pm20}$ state and the qudit, the global state becomes
\begin{equation}
\mu_1\mu_0\ket{-2}\ket{-2}+\mu_1\nu_0\ket{2}\ket{-2}+\nu_1\mu_0\ket{2}\ket{2}+\nu_1\nu_0\ket{2}\ket{2}
\end{equation}
which is, exactly, 
\begin{equation}
(\mu_1\ket{-2}+\nu_1\ket{2})\otimes(\mu_0\ket{-2}+\nu_0\ket{2}).
\end{equation}
The combination that erases the remains of the extracted qubit is performed with a probability over the 97.60\%, for an $m=1$ adder. At this point, all the original qubits are again separated. 

Figure \ref{example} summarizes the whole procedure in the XY plane. The multiplexed qudit can be in any superposition of a certain number of different coherent states. The set of the possible coherent states is reduced after each extraction. The qubit value is recovered by identifying logical $\0_L$ and $\1_L$ with two different sets of states that, after the qubit has been recovered, interfere to produce a superposition in the new smaller qudit space. 

It is also interesting to see the effects of the addition of a new user after the system has been designed. Imagine that, once an upper limit around $\alpha=222$ has been defined, a new user wants to enter the channel. Its qubit $\ket{\psi_3}=\mu_3\ket{-2}+\nu_3\ket{2}$ can be multiplied by $M'=\sqrt{10}$, resulting in $\mu_3\ket{-\sqrt{40}}+\nu_3\ket{\sqrt{40}}$. The whole procedure could be repeated reserving the second channel for this new user. The extraction of qubit $\ket{\psi_2}$ would be slightly impaired. Now the most separated state from $\ket{2}$ would be $\ket{2.2832}$ and the probability of correction goes down from 97.60\% to 96\%. The probability of correct erasure at the combiner similarly reduces from the 97.56\% to the 97.42\%, for $m=3$, and from the 67.66\% to the 66.15\% for $m=2.4$. In the next steps, erasure in the combiner is not affected much by the existence of an additional user. More dangerous is the decoding of the new state. The $\ket{\psi_0}$ and $\ket{\psi_3}$ states are combined as 
\begin{eqnarray}
\nonumber\mu_3\mu_0\ket{-8.3246}\ket{-2.6325}&+&\mu_3\nu_0\ket{-4.3246}\ket{-1.3675}\\
+\hspace{5ex}\nu_3\mu_0\ket{4.3246}\ket{1.3675}&+&\nu_3\nu_0\ket{8.3246}\ket{2.6325}
\end{eqnarray}
after the division by $M'$. Now, $\epsilon\approx 0.3162$ in the worst case and the probability of success is reduced to only the 81.87\%. 

With this example, it becomes clear that the new user slots should be filled first from the lower amplitude states. If an error occurs, at least the upper channels will have been transmitted correctly.

\section{Discussion}
\label{Conclusion}
We have presented a quantum version of the CDMA multiple access technique that can be used to send $N$ coherent state qubits together as a superposition of $2^N$ coherent states. Different channels are encoded at different levels of detail in the amplitude of the transmitted state, so that all the other channels can be taken as noise. The users employ encodings that are separated by factors of a parameter $M$. The methods of coherent state qubit error correction by means of teleportation allow us to separate the users at the receiver. User information must be extracted sequentially as a consequence of the strong correlations between the qubits that form the transmitted qudit. New users can be easily added to the channel, even above first design limits, at the cost of smaller success rates. The basic resource are entangled cat states.

Coherent state multiplexing permits to send the quantum data of many users with a single transmission. Instead of sending each qubit at different times, or at different frequencies, so that they do not interfere, we can make a better use of the channel. Apart from this advantage, we have to consider the effects of decoherence on the new encoding. Transmission through a real medium, such as a lossy optical fibre, can destroy the coherence of the quantum data. In particular, superpositions of coherent states can be very sensitive to decoherence and show a decay rate which is proportional to their separation in the XY plane \cite{WM94}. This imposes a limit on the maximum amplitude of our multiplexed states. 

Nevertheless, coherent state multiplexing can also have some benefits. The proposed scheme gives a partial protection against decoherence. The two most important effects of decoherence during the transmission of superpositions of coherent states of a real $\alpha$ are amplitude reduction and $\pi$ phase shifts, or Z errors \cite{GVR04}. The effect of the channel on the transmission of such states can be modelled as a beamsplitter in which the superposition is mixed with the vacuum. The transmission coefficient of the beamsplitter is determined by the channel losses. The situation is equivalent to the problem of erasure we found in Section \ref{adder} when we discussed the adder. 

In this model, amplitude reduction is determined by the part of the field that leaks into the medium. The teleportation extraction of the demultiplexer provides an effective regeneration. In the same way that the extraction of a qubit cleans the traces of the other channels, it will also correct the presence of noise or amplitude reductions of the same order of amplitude as the discarded data. Z errors are also related to the field that the loss beamsplitter extracts from the channel. As we have discussed for the adder, if this field is measured and an odd number of photons is found, there will be a sign shift between the terms with a positive and a negative amplitude. In the case of decoherence, we cannot know whether this Z error has occurred or not. But, as it happened in the adder, this error will only affect the leading term. The sign of the amplitude depends only on the qubit of order $N-1$. All the other qubits are protected by the highest amplitude qubit, which will take all the Z errors. After the extraction, the new leading term will be exposed to Z errors. However, in many scenarios, the multiplexed state will travel most of the distance and decoherence between the extractions at the location of each user will probably be less important. With coherent state multiplexing, we only have to worry for the dephasing of one qubit. 

In the design of the system there are some tradeoffs that need to be taken into account. On one hand, large cats should be avoided, as coherent states of high intensity are challenging to obtain. Furthermore, while individual coherent states are relatively robust against losses, large cat states are very sensitive to loss and inefficient detection \cite{GV08}. From that point of view, the maximum amplitude value of $M^{N-1}\alpha$ should be kept small. On the other hand, it is desirable to minimize errors in the extraction. Efficiency is limited by the error correction technique, which succeeds with probability $e^\frac{-|\alpha|^2|\epsilon|^2}{2}$. $\epsilon$ has been shown to be of the order of $\frac{1}{M}$. Extraction will be more efficient for values of $M\gg \alpha$. Additionally, $\alpha$ should be high in order to easily distinguish between $\0_L=\alm$ and $\1_L=\al$ logical values of the qubit. From that point of view, large values of $\alpha$ and $M$ are preferred. 

Future work must be done to carefully analyse the election of these parameters. In particular, it would be interesting to find balanced error probabilities. The given examples present almost null qubit confusion error probabilities, but appreciable error correction failure probabilities in qubit extraction. An optimum balance between these values needs to be found so that the limiting probability factor can be raised. Adaptive multiplexing, where each channel is multiplied by a different value $M_k$, which needs not to be the power of a fixed value, can also help to take the most out of the existing coherent state technology. Other adaptations of the model that could be explored are multiplexers with squeezed states or data encodings that make full use of the available state space instead of confining the data to the X quadrature. 

Among the next tasks to be done in coherent multiplexing, it is an exhaustive calculation of the fidelity of the received qubits. In this paper, we have proved such a multiplexing is possible, but detailed numerical analysis of fidelity and behaviour against photon loss and other imperfections is left for another occasion. It would also be worth trying to build a simple experimental prototype, for instance for a Quantum Key Distribution application.  

The presented multiplexing technique can be an interesting addition to the coherent state quantum information model. In particular, it is a useful complement to fibre transmission of quantum information in coherent states. With the present technology two or three qubits could be sent together with a reasonable probability of success. Further advances will be constrained to the creation of larger entangled coherent state resources. 

\section*{Acknowledgements} 
J.C. Garc\'ia-Escart\'in would like to thank Barry Sanders and Aggie Bra\'nczyk for interesting coffee-break conversations and advice at QTS5 and EYSCQI in Valladolid and Vienna. This work has been funded by MEC and FEDER project Ref. TEC2007-67429-C02, and by JCyL, Grant No. VA001A08. 

\newcommand{\noopsort}[1]{} \newcommand{\printfirst}[2]{#1}
  \newcommand{\singleletter}[1]{#1} \newcommand{\switchargs}[2]{#2#1}

\end{document}